\documentclass{article}
\usepackage[margin=2cm]{geometry} 
\usepackage{authblk} 
\usepackage{graphicx} 
\usepackage{multicol} 
\usepackage{biblatex} 
\usepackage{subcaption} 
\usepackage{physics} 
\usepackage{comment} 
\usepackage{float}

\title{Robust Wada Boundaries and Entropy Scaling in pp-Wave Spacetimes}
\author[1,*]{Pedro Henrique Barboza Rossetto}
\author[2]{Vanessa Carvalho de Andrade}
\author[2]{Daniel Müller}
\affil[1]{Escola de Artes, Ciências e Humanidades, Universidade de São Paulo \\ Rua Arlindo Bettio, CEP 03828-000, 1000 São Paulo, SP, Brazil}
\affil[2]{Instituto de Física, Universidade de Brasília \\
70919-970 Brasília, DF,  Brazil}
\affil[*]{Corresponding author: phbrossetto@gmail.com}  
\date{}

\newcommand{\paren}[1]{\left(#1\right)}
\newcommand{\colch}[1]{\left[#1\right]}

\begin{document}
\maketitle
\begin{abstract}
    We study the dynamics of the geodesics of pp-wave spacetimes with polynomial profiles, which are dynamically equivalent to the motion of a classical particle in a two-dimensional harmonic polynomial potential. We demonstrate that the Wada property of the escape basins is robust under variation of the polynomial degree, i.e., the basin boundaries remain maximally intermingled as the number of escape channels increases. We further provide a quantitative characterization of the degree of dynamical uncertainty by computing the basin entropy $S_{b}$ and the boundary basin entropy $S_{bb}$. We find that these measures increase monotonically with the polynomial degree, indicating enhanced unpredictability of the final state of the system. We also show that $S_{bb}$ is greater than $\ln(2)$ for $n>3$, and this confirms that the basin boundaries are fractal.

\hphantom{line break}

\end{abstract}

\begin{multicols}{2}
\section{Introduction}
Since the first direct detections of gravitational waves, exact radiative spacetimes have provided a natural arena for investigating nonlinear effects and limits of predictability in general relativity. Within this class, plane-fronted waves with parallel rays (pp-waves) play a prominent role. Originally introduced by Brinkmann, these solutions admit a covariantly constant null Killing vector and satisfy harmonic constraints on the metric profile, which enable a systematic and tractable analysis of geodesic motion \cite{brinkmann1923riemann, stephani2003exact}. When the profile function is independent of the null coordinate, the transverse geodesic dynamics can be mapped onto an effective two-dimensional Hamiltonian system governed by a harmonic potential. This correspondence allows well-established methods from nonlinear dynamics and chaotic scattering to be applied in a relativistic setting \cite{podolsky1998chaos, podolsky1998chaotic, podolsky1999smearing}.

Chaotic behavior in non-homogeneous pp-wave spacetimes was rigorously demonstrated by \citeauthor{podolsky1998chaos} \citeyear{podolsky1998chaos}, who showed that distinct asymptotic escape channels are separated by fractal basin boundaries and that the final state displays extreme sensitivity to initial conditions \cite{podolsky1998chaos}. From a dynamical perspective, this places the reduced geodesic dynamics within the broader class of open Hamiltonian systems. A paradigmatic example in this context is the Hénon–Heiles model, for which, above the escape energy, multiple exits coexist and the corresponding exit basins exhibit fractal boundaries \cite{henon1964applicability,rod1973pathology,aguirre2001wada}. In a number of such systems, basin boundaries satisfy the more restrictive Wada property, whereby every boundary point is simultaneously shared by all basins \cite{kennedy1991basins,aguirre2001wada}. The presence of Wada boundaries represents a stronger form of unpredictability in the geometric sense of basin structure, since arbitrarily small uncertainties in initial conditions cannot be associated with any proper subset of final outcomes \cite{kennedy1991basins,aguirre2001wada,poon1996wada}.

The verification of the Wada property has traditionally relied on invariant-manifold arguments, such as the existence of accessible unstable periodic orbits whose invariant manifolds intersect all basins. More recently, a set of numerical criteria has been developed, enabling systematic and reproducible Wada checks in high-resolution phase-space representations. Among these, the grid method identifies Wada points through successive refinements of discretized neighborhoods near basin boundaries, providing a systematic and reproducible criterion for the verification of the Wada property in numerical phase-space representations \cite{kennedy1991basins,aguirre2001wada,daza2015testing}.

In parallel with such geometric diagnostics of basin structure, quantitative measures of uncertainty associated with multi-exit systems have been introduced. In particular, basin entropy quantifies the local mixing of different outcomes within a finite-resolution covering of phase space and allows direct comparisons between systems with different numbers of exits. The closely related boundary basin entropy focuses exclusively on boundary regions and provides a sufficient condition for fractality: values exceeding ln 2 indicate fractal basin boundaries even when the analysis is performed at a fixed box size $\varepsilon$ \cite{daza2016basin}.

These concepts have been shown to be directly applicable to relativistic systems. In particular, \citeauthor{rossetto2020wada} \citeyear{rossetto2020wada} provided numerical evidence that geodesic escape basins in pp-wave spacetimes with cubic harmonic profiles possess Wada boundaries, establishing a direct link between exact gravitational wave solutions and classical chaotic scattering phenomena \cite{rossetto2020wada}. This result motivates the investigation of how robust the Wada property is under variations of the polynomial degree of the wave profile and how the associated uncertainty evolves as the number of escape channels increases.

By varying the polynomial degree, one obtains a controlled family of open Hamiltonian systems with an increasing number of escape channels and progressively more intricate phase-space structure. This provides a natural setting to investigate both the robustness of Wada basin topology and the scaling of dynamical uncertainty. The analysis focuses on the identification of escape channels, the numerical verification of the Wada property using the grid method, and the quantification of uncertainty in the final state through basin entropy and boundary basin entropy. The paper is organized as follows. Section 2 reviews the formulation of pp-wave spacetimes and their dynamical equivalence to two-dimensional harmonic polynomial potentials. Section 3 presents the construction of escape basins in position and momentum space and discusses their geometric properties. Section 4 verifies the Wada property using the grid method and relates the results to classical concepts from dynamical systems. Section 5 introduces basin entropy as a quantitative measure of uncertainty and analyzes its dependence on the polynomial degree. Finally, Section 6 discusses the implications of the results.

\section{pp-Wave Spacetimes}

Plane-parallel gravitational waves, or simply pp-waves, are a class of exact solutions to Einstein's Field Equations. We can write the metric using Brinkman coordinates $(x,y,z,u,v)$
\begin{equation}
    \label{pp_wave_metric}
    \dd{s}^2=\dd{x}^2 + \dd{y}^2 - 2\dd{u}\dd{v} - H(x,y,u)\dd{u}^2.
\end{equation}
The only non-zero component of the Ricci tensor is
\begin{equation}
    \label{ricci_ten_pp_uu}
    R_{uu}=\frac{1}{2}\paren{\pdv[2]{H}{x}+\pdv[2]{H}{y}}.
\end{equation}
Therefore, for the metric \eqref{pp_wave_metric} to satisfy the vacuum Einstein's Field Equations ($R_{ab}=0$), we simply need that
\begin{equation}
    \label{H_condition}
    \nabla^{2}_{x,y}H = \pdv[2]{H}{x}+\pdv[2]{H}{y} = 0.
\end{equation}

The geodesic equations for this metric are
\begin{align}
    \label{geod1}
    \ddot{x} + \frac{1}{2}\pdv{H}{x}\dot{u}^2 = 0,
    \\
    \label{geod2}
    \ddot{y} + \frac{1}{2}\pdv{H}{y}\dot{u}^2 = 0,
    \\
    \label{geod3}
    \ddot{u} = 0,
    \\
    \label{geod4}
    \ddot{v} + \frac{1}{2}\pdv{H}{x}\dot{x}\dot{u}
    + \frac{1}{2}\pdv{H}{y}\dot{y}\dot{u} 
    + \frac{1}{2}\pdv{H}{u}\dot{u}^2 = 0,
\end{align}
where the dot represents the total derivative with respect to the geodesic parameter $s$. Equation \eqref{geod3} can be solved analytically, and we obtain an afine relation between the coordinate $u$ and $s$. We then re-parametrise the geodesic using the coordinate $u$ as the parameter, and we obtain
\begin{align}
    \label{geodx}
    \ddot{x} + \frac{1}{2}\pdv{H}{x} = 0,
    \\
    \label{geody}
    \ddot{y} + \frac{1}{2}\pdv{H}{y} = 0,
    \\
    \label{geodv}
    \ddot{v} + \frac{1}{2}\pdv{H}{x}\dot{x}
    + \frac{1}{2}\pdv{H}{y}\dot{y} 
    + \frac{1}{2}\pdv{H}{u} = 0.
\end{align}
Notice that the equations for $x$ and $y$ are decoupled from the equation for $v$; therefore, we only need to solve equations \eqref{geodx} and \eqref{geody} to obtain the motion on the $x,y$ plane. These two equations of motion can also be derived from the classic Hamiltonian $\mathcal{H}$ given by
\begin{equation}
    \label{hamiltonian}
    \mathcal{H} = \frac{p_x^2+p_y^2}{2}+V(x,y,u),
\end{equation}
where,
\begin{equation}
    \label{pot_as_H}
    V(x,y,u) = \frac{1}{2}H(x,y,u).
\end{equation}

The $u$ dependence of the function $H(x,y,u)$ creates a wave pulse, which ``smears'' the trajectory of the particles \cite{podolsky1999smearing}.  For that reason, in the present work, we choose $H(x,y,u)$ to be independent of $u$. From equation \eqref{pot_as_H} we see that the potential of the system also has to satisfy the harmonic condition \eqref{H_condition}. Therefore, the problem of geodesic equations in the $x,y$ plane in pp-wave spacetimes is equivalent to the classical problem of the motion of a particle in a two-dimensional harmonic potential. In this paper, we consider the case where the potential $V=V_n(x,y)$ is a harmonic polynomial of degree $n$. These polynomials can be found by taking the real (or imaginary) part of the binomial expansion of $(x+iy)^n$ \cite{brown2009complex}, thus, we define
\begin{equation}
    \label{potential}
    V_n(x,y) = \frac{1}{2}\Re\colch{(x+iy)^n}.
\end{equation}
It is worth noting that if one takes the imaginary part of the binomial expansion, one would not obtain a dynamically unique system. The new system would simply be a $2\pi/n$ clockwise rotation of the system described by \eqref{potential} \cite{brown2009complex}.

In a convenient alternative formulation, we could express all the equations of motion in terms of angular coordinates $r$ and $\theta$. In this case, the potential \eqref{potential} is simply
\begin{equation}
    V_n = \frac{1}{2}r^n\cos(n\theta),
\end{equation}
and the geodesic equations become
\begin{align}
    &\ddot{r}-r\dot{\theta}^2+\frac{1}{2}nr^{n-1}\cos(n\theta) = 0,
    \\
    &\ddot{\theta} + 2\frac{\dot{r}\dot{\theta}}{r} - \frac{1}{2}nr^{n-2}\sin(n\theta) = 0.
\end{align}
These equations are manifestly simpler than the ones in Cartesian coordinates. Nevertheless, polar coordinates introduce a coordinate singularity at $r=0$ and, hence, hinder the numerical solutions near the origin. For this reason, we solve the equations of motion of the system in Cartesian coordinates. In Appendix \ref{app_numerics}, we discuss our numeric implementation of the geodesic problem.
\section{Escape Basins}
\label{sec:basins}
\begin{figure*}
    \centering
    \begin{subfigure}{0.45\textwidth}
        \includegraphics[width=\textwidth]{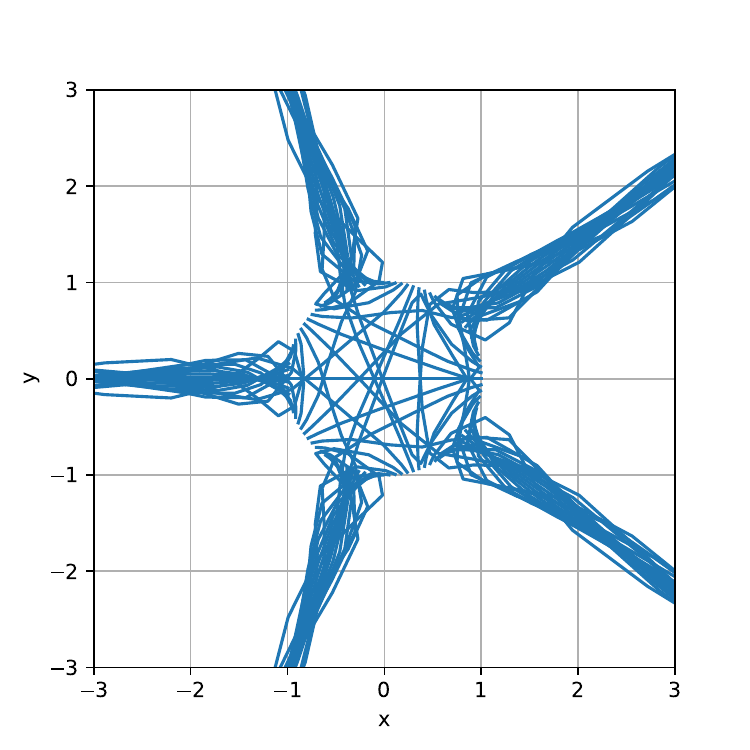}
        \caption{}
        \label{fig:trajectories_n=5}
    \end{subfigure}
    \begin{subfigure}{0.45\textwidth}
        \includegraphics[width=\textwidth]{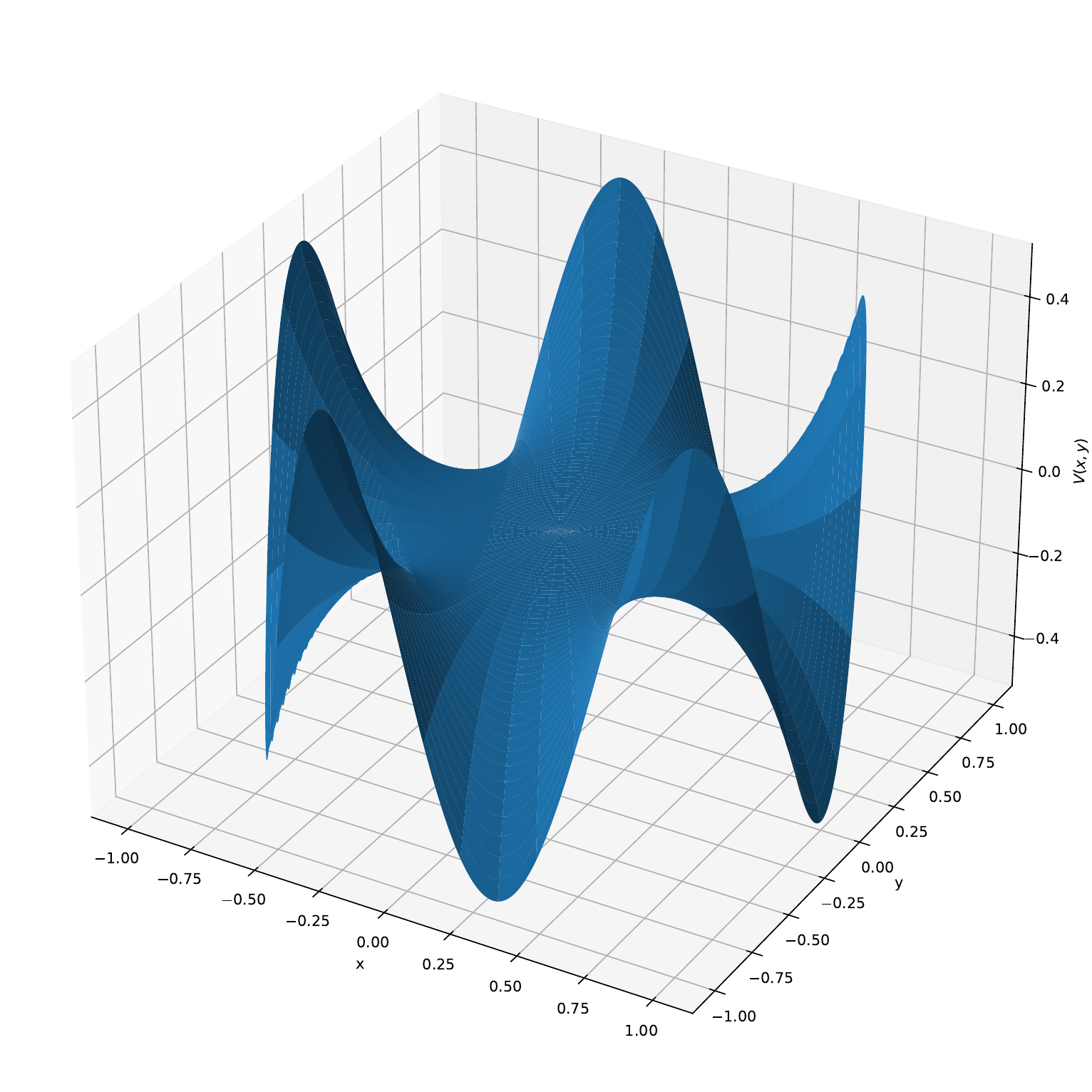}
        \caption{}
        \label{fig:potential_n=5}
    \end{subfigure}
    \caption{(a) Trajectory of a ring of particles equally spaced around the circle $x^2+y^2=1$, initially at rest. (b) Potential \eqref{potential} for $n=5$.}
    \label{fig:traj_and_pot}
\end{figure*}

Free particles in pp-wave spacetimes, following the equations \eqref{geodx}, \eqref{geody}, \eqref{pot_as_H} and
\eqref{potential}, escape to infinity (in the x-y plane) by the regions where the potential tends to $-\infty$ \cite{rod1973pathology,podolsky1998chaos,podolsky1998chaotic,rossetto2020wada}. Areas of positive, infinite potential separate these escape regions, and, therefore, they are distinct and well-defined \textit{escape channels}. For a harmonic polynomial potential of degree $n$, the sector angle $\alpha$ of the escape channels is simply given by $\alpha=2\pi/n$. Therefore, there are $n$ escape channels. Considering that a particle has angle $\theta$ at $x$-$y$ spatial infinity, we say that it has escaped through the channel $c=\text{ceil}(\theta/\alpha)$, where $\text{ceil}$ is the ceiling function. This behavior is exemplified in Figure \ref{fig:traj_and_pot}, where we plot the trajectory of a ring of particles initially at rest (Figure \ref{fig:trajectories_n=5}) for the $n = 5$ potential (Figure \ref{fig:potential_n=5}).

Escape basins are a standard plot when it comes to chaotic scattering \cite{poon1996wada,daza2018wada,desouzafilho2020fractal, aguirre2001wada,podolsky1998chaotic}. These graphs relate the initial conditions of a dynamical system with the escape channel to infinity that the particle took. The escape basins for the problem of geodesic motion in pp-waves spacetime are known to be fractal in the literature \cite{podolsky1998chaotic,podolsky1998chaos}. In this section, we reproduce the choice of one-dimensional and two-dimensional escape basins for this system, using the numerical methods discussed in Appendix \ref{app_numerics}.

\begin{figure*}
    \centering
    \includegraphics[width=0.6\linewidth]{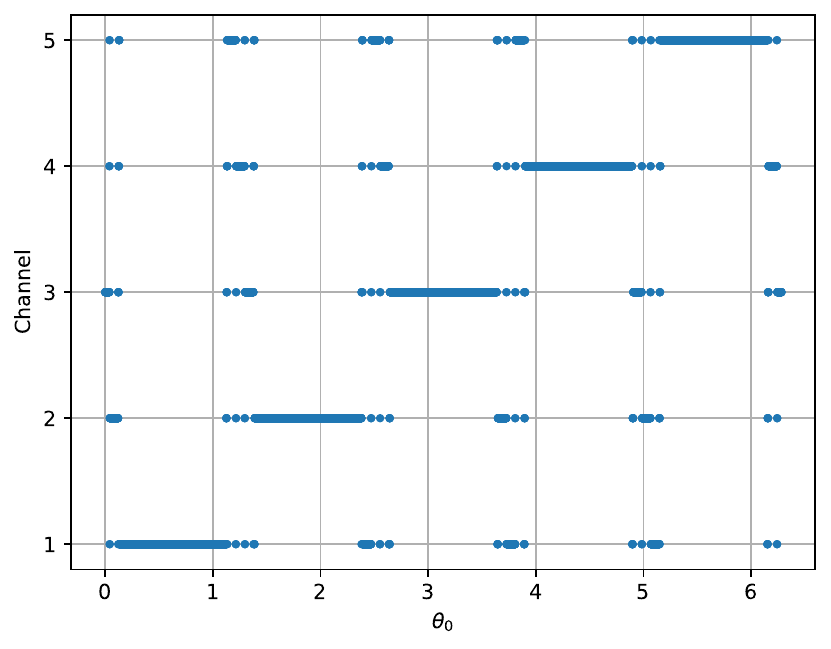}
    \caption{One-dimensional escape basin for $n=5$ and for particles distributed along a unit circle with zero initial velocity.}
    \label{fig:basin_n=5_1D}
\end{figure*}

A possible choice of a one-dimensional family of initial conditions is $x(0) = \cos(\theta_0)$, $y(0) = \sin(\theta_0)$ and $p_x(0) = p_y(0) = 0$, with the parameter $\theta_0$ varying between $0$ and $2\pi$. The escape basin for this choice, with $n=5$, is shown in Figure \ref{fig:basin_n=5_1D}, which reproduces Figure 11 of \cite{podolsky1998chaotic}. From Figure \ref{fig:basin_n=5_1D}, one can notice that there are ``lengthy steps of certainty'' where if we choose a nearby initial condition $\theta_0$, the escape channel will remain the same. Nevertheless, the boundary between these steps shows fractality, as was already discussed in the literature \cite{podolsky1998chaos, podolsky1998chaotic}. Moreover, every boundary is not only fractal, but it also seems to be the boundary of all regions. This is the Wada property, which is discussed and verified in Section \ref{sec:wada}. 

\begin{figure*}
    \centering
    \begin{subfigure}{0.45\textwidth}
        \includegraphics[width=\textwidth]{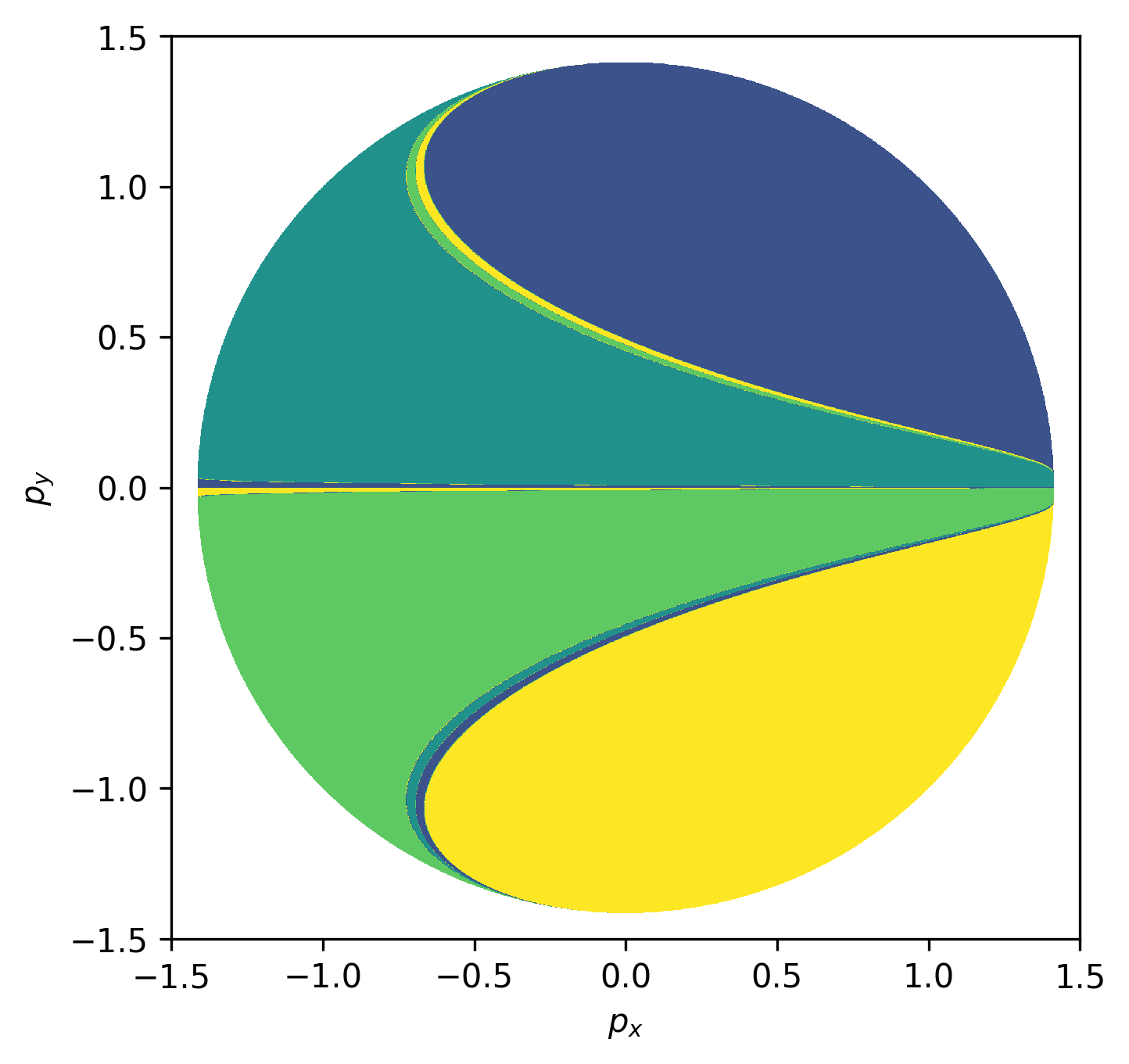}
        \caption{}
        \label{fig:basin_n=4}
    \end{subfigure}
    \begin{subfigure}{0.45\textwidth}
        \includegraphics[width=\textwidth]{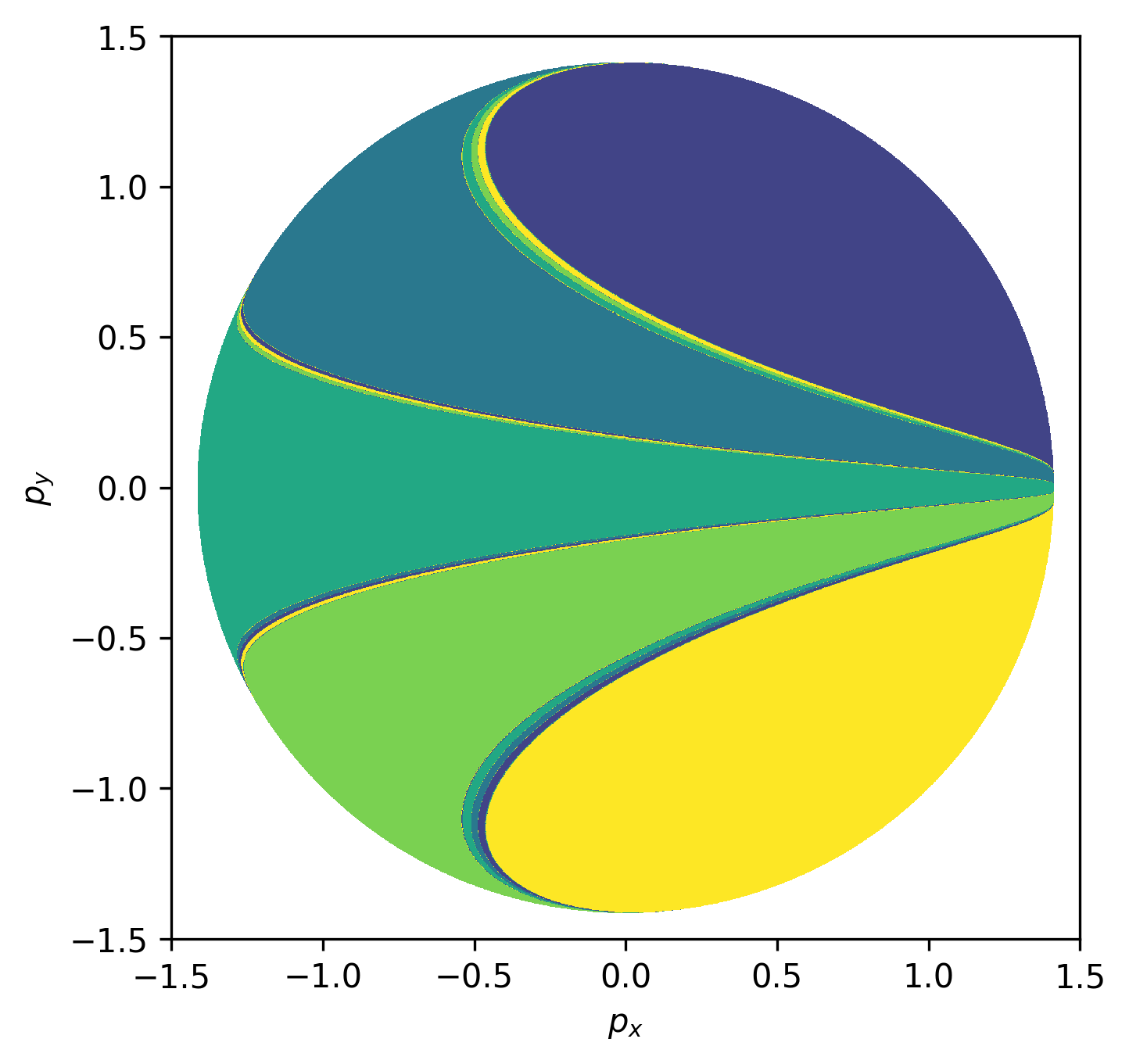}
        \caption{}
        \label{fig:basin_n=5}
    \end{subfigure}
    \caption{Exit basins for the $p_x$-$p_y$ plane for the polynomial degree values of (a) $n=4$ and (b) $n=5$. The region outside of the disk leads to unphysical initial conditions. The color gradient indicates the escape channels, with dark blue being the first channel and yellow being the last channel.}
    \label{fig:exit_basins}
\end{figure*}

\begin{figure*}
    \centering
    \includegraphics[width=0.9 \textwidth]{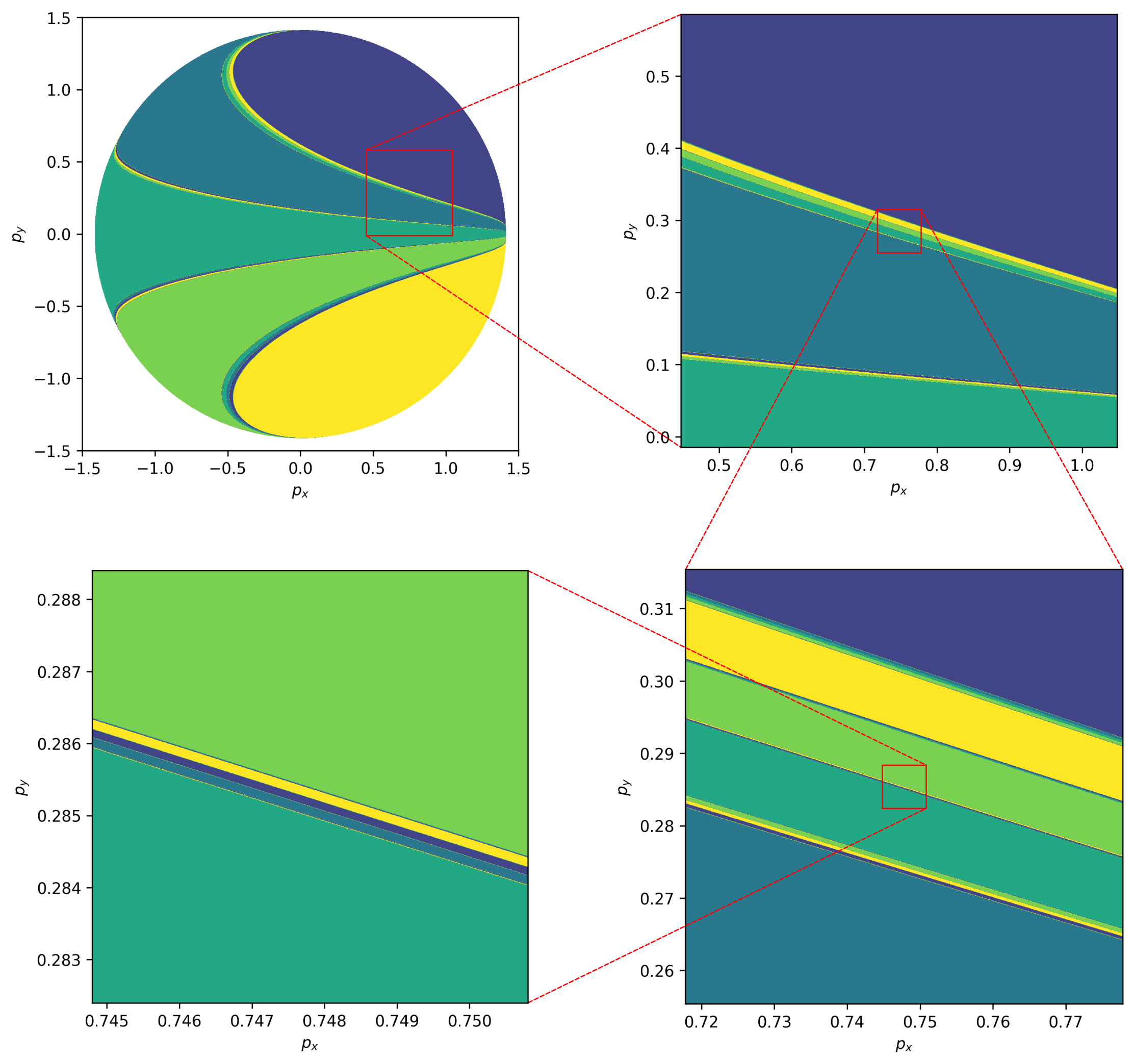}
    \caption{Three zoom levels of the exit basin for $n=5$. Each successive zoom level gives a magnification of $10$ times. The color gradient indicates the escape channels, with dark blue being the first channel and yellow being the fifth channel.}
    \label{fig:basin_n=5_zoom}
\end{figure*}

Other escape basins can be constructed for the choice of initial conditions in the momentum plane. In order to compare our results with the previous literature \cite{podolsky1998chaotic,podolsky1998chaos}, we make the following choices for the $p_x-p_y$ escape basin: (i) $p_x$ and $p_y$ vary between $-1.5$ and $1.5$, (ii) $y=0$, and (iii) $x = (2E - p_x^2 - p_y^2)^{1/n}$, where $E=1$ is the energy of the particle. Using this choice, the basins for $n=4$ and $n=5$ are shown in Figure \ref{fig:exit_basins}. The boundary structures of the basin for $n=5$ are shown in Figure \ref{fig:basin_n=5_zoom}. In this figure, the fractal structure becomes visually apparent, as well as the Wada property.

\section{Wada Boundaries}
\label{sec:wada}
The Wada property was first introduced as a concept in topology \cite{yoneyama1917theory}. It was considered a system with three regions in such a way that any boundary point was a boundary of all three regions. Such a system is said to possess the Wada property. Apart from being an interesting mathematical property, it was found that many physical systems also present the Wada property \cite{aguirre2001wada,desouzafilho2020fractal,bautista2021chaotic,fernandez2022weak}. In the context of chaotic dynamics, this property increases the unpredictability of the system since small variations in the initial condition can make the system evolve to all its different final states. Examples where the Wada property was found include: the Hénon-Heiles system \cite{aguirre2001wada}, light scattering in a pair of black holes \cite{desouzafilho2020fractal}, charged particle scattering in a weakly magnetized black hole \cite{bautista2021chaotic}, three-dimensional scattering \cite{fernandez2022weak}, among others.

It is known in the literature that the escape basin of geodesics in pp-wave spacetimes is fractal \cite{rod1973pathology,podolsky1998chaos,podolsky1998chaotic}. It is also known that in a restricted case --- $n=3$ and in one angular dimension --- the system possesses the Wada property in the \cite{rossetto2020wada}. In this section, we demonstrate numerically that the Wada property of the escape basins of the system is robust. That is, the Wada property is verified for: both the one and two-dimensional families of escape basins (as discussed in Section \ref{sec:basins}), and for increasing values of $n$. We use the method presented by \citeauthor{daza2015testing} in \citeyear{daza2015testing} \cite{daza2015testing} to determine if our system possesses the Wada property. We refer to this method hereafter as the grid method. It consists of taking a basin, such as the one presented in Figure \ref{fig:exit_basins}, dividing it into a grid, and then classifying each point in the grid according to its neighbors. Interior points are points that are completely surrounded by points of the same color. Boundary points are points that have at least one neighbor of a different color. A boundary point is called a Wada boundary point if it has neighboring points of all colors.

At first, the grid method loads the basin and classifies all points according to their neighbors. However, this initial classification can incorrectly classify a Wada point as a non-Wada point due to insufficient resolution of the loaded basin. To resolve this issue, we refine the grid near the non-Wada boundary points and verify if there are no other colors near them. If other colors are found, the original point is ``promoted'' as a boundary point of more regions. At each refinement, the method doubles the number of points checked near the boundary point in question. The refinements stop when a certain stability criterion is met; for more details on the stop condition, see Appendix \ref{app_numerics}.

Let $G_i^q$ be the set of boundary points of $i$ regions at the iteration (refinement) $q$ of the code. Then, we define the ratios $W_i^q$ as
\begin{equation}
    \label{W_def}
    W_i^q = \frac{|G_i^q|}{\sum_{i=2}^n |G_i^q|},
\end{equation}
where $|\cdot|$ is the cardinality operator of sets and $n$ is the total number of regions. In our problem, $n$ is given by the degree of the polynomial, as displayed in equation \eqref{potential}, which justifies the use of the same variable name for both quantities. Given this definition of the parameter $W$, the grid method ascertains that a system possesses the Wada property if $W_n^{q_{max}}$ is sufficiently close to $1$ at the last iteration $q_{max}$.

\begin{figure*}
    \centering
    \begin{subfigure}{0.45\textwidth}
        \includegraphics[width=\textwidth]{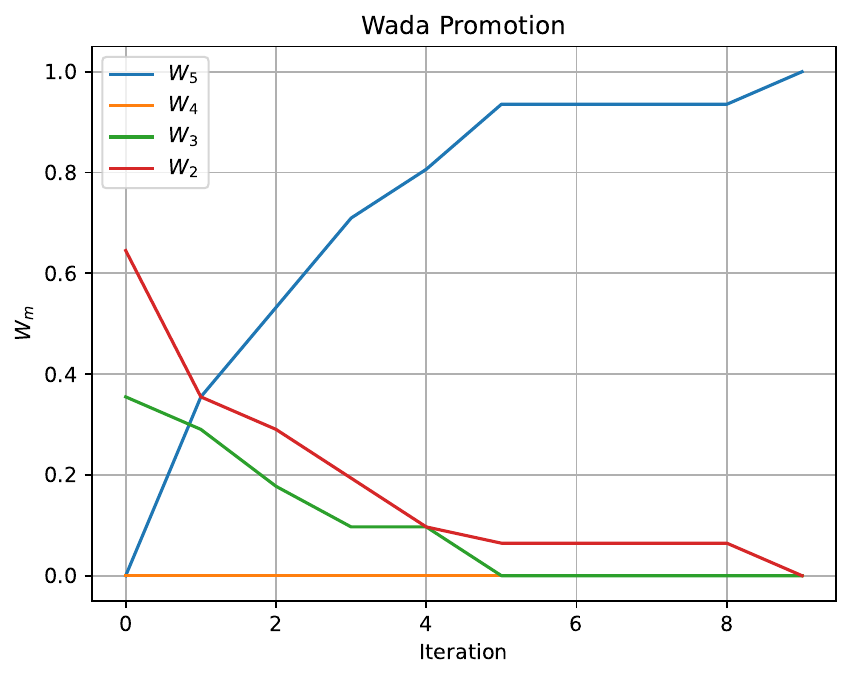}
        \caption{}
        \label{fig:grid_wada_n=4}
    \end{subfigure}
    \begin{subfigure}{0.45\textwidth}
        \includegraphics[width=\textwidth]{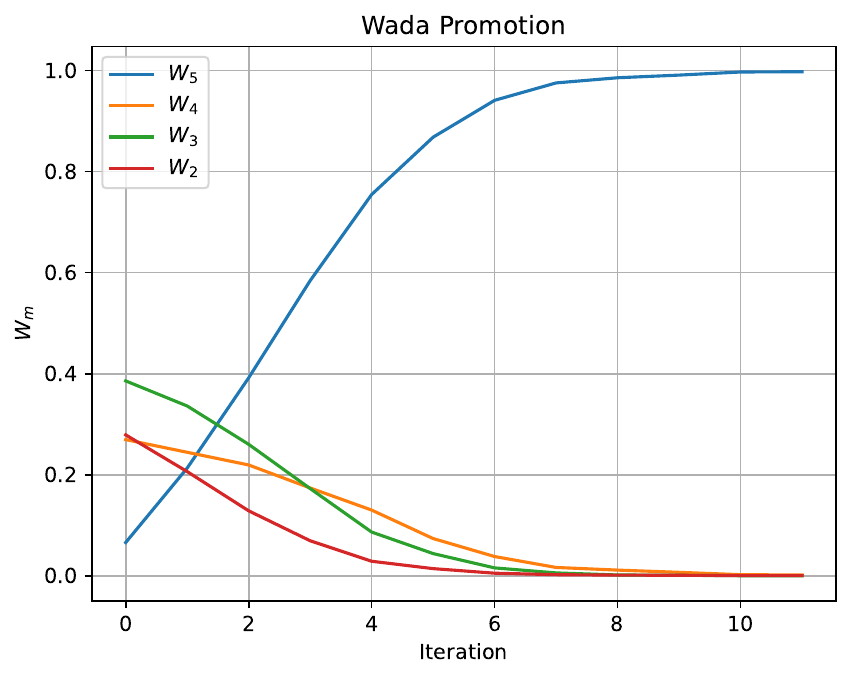}
        \caption{}
        \label{fig:grid_wada_n=5}
    \end{subfigure}
    \caption{Grid method applied to the $n=5$ basins. (a) For the angular position space, displayed in Figure \ref{fig:basin_n=5_1D}, and (b) for the momentum plane basin, displayed in Figure \ref{fig:grid_wada}(b). For both boundaries, the grid method concludes that the boundaries have the Wada property.}
    \label{fig:grid_wada}
\end{figure*}

Figure \ref{fig:grid_wada} shows the evolution of all $W_i$, per iterations $q$ of the grid method, for the $n=5$ basins (see figures \ref{fig:basin_n=5_1D} and \ref{fig:exit_basins}(b)). Both basins have the Wada property. We also checked the basins for $n$ values up to $10$ and the values of $W_n^{q_{max}}$ are noted in Table \ref{tab:wada_per_n_1D}, for the one-dimensional basins, and in Table \ref{tab:wada_per_n_2D} for the two-dimensional ones. Within numerical precision, we see that all basins tested have the Wada property, regardless of $n$ and regardless of which set of initial conditions is chosen. 

\begin{table}[H]
\centering
\begin{tabular}{|c|c|}
\hline
$n$  & $W_n^{q_{max}}$ \\ \hline
$3$  & 1.00            \\ \hline
$4$  & 1.00            \\ \hline
$5$  & 1.00            \\ \hline
$6$  & 1.00            \\ \hline
$7$  & 1.00            \\ \hline
$8$  & 1.00            \\ \hline
$9$  & 0.91            \\ \hline
$10$ & 1.00            \\ \hline
\end{tabular}
\caption{Final value of $W_n$ for different values of $n$ for the one-dimensional basin in angular space.}
\label{tab:wada_per_n_1D}
\end{table}

\begin{table}[H]
\centering
\begin{tabular}{|c|c|}
\hline
$n$  & $W_n^{q_{max}}$ \\ \hline
$3$  & 0.98958        \\ \hline
$4$  & 0.99718        \\ \hline
$5$  & 0.99737        \\ \hline
$6$  & 0.99737        \\ \hline
$7$  & 0.99920        \\ \hline
$8$  & 1.00000        \\ \hline
$9$  & 0.99939        \\ \hline
$10$ & 0.99986        \\ \hline
\end{tabular}
\caption{Final value of $W_n$ for different values of $n$ for the two-dimensional basin in momentum space.}
\label{tab:wada_per_n_2D}
\end{table}

\section{Basin Entropy}
In this section, we analyze the basin entropy of the basins of escape of the polynomial pp-waves with degree $n$. The basin entropy is a quantitative tool that measures the degree of uncertainty of a dynamical system. This tool was introduced by \citeauthor{daza2016basin} in \citeyear{daza2016basin} \cite{daza2016basin}. For a full discussion of the definition, properties, and applications of this entropy, the reader is referred to their paper. In the following, we introduce the main definitions so that this paper's results can be presented and discussed.

Considering an escape basin, e.g. Figure \ref{fig:exit_basins}, we can choose a cover $\Omega$ for it. In the basins of this paper, we consider the cover to be a set of $N$ non-overlapping square boxes with side length $\varepsilon$, in a way that each box contains several points inside. The total entropy $S$ is defined to be
\begin{equation}
    \label{entropy}
    S = \sum_{i=1}^{N}\sum_{j=1}^{m_i}p_{i,j}\log(\frac{1}{p_{i,j}}),
\end{equation}
where $i$ is the box index, $m_i$ is the total number of colors in box $i$, and  $p_{i,j}$ is the probability of finding the color $j$ in box $i$. Then, the \textit{basin entropy} $S_b$ is defined to be $S/N$. It can be shown \cite{daza2016basin} that $S_b$ can also be written as
\begin{equation}
    \label{S_b}
    S_b = \sum_{k=1}^{k_{\text max}}A_k\varepsilon^{\alpha_k}\log(m_k),
\end{equation}
where $k$ is a label for the different basin boundary (up to $k_{\text max}$), $A_k$ is a normalization constant that depends on the number of boxes in the boundary, $\alpha_k$ is the uncertainty exponent \cite{grebogi1983final}, and $m_k$ is the number of colors in the $k$th boundary. For systems with the Wada property, all boundaries between any number of colors are also boundaries of every color, and every boundary box contains all the colors in it. Therefore, for Wada systems, equation \eqref{S_b} simplifies to
\begin{equation}
    \label{S_b-wada}
    S_b = A\varepsilon^\alpha\log(N_A).
\end{equation}

As it was shown in Section \ref{sec:wada}, the escape basins of the pp-wave system of degree $n$ possess the Wada property, therefore, equation \eqref{S_b-wada} is applicable. By increasing the degree $n$ of the potential \eqref{potential}, we can expect that the entropy will increase both due to the factor $\log(N_A)$ and the factor $A$. The increase of $n$ also leads to an increase in the intricacy of the boundary, causing a higher uncertainty of the final state of the system. Therefore, in the language of uncertainty exponent, we expect $\alpha$ to decrease as $n$ increases \cite{grebogi1983final}. 

\begin{figure*}
    \centering
    \begin{subfigure}{0.51\textwidth}
        \includegraphics[height=0.22\textheight]{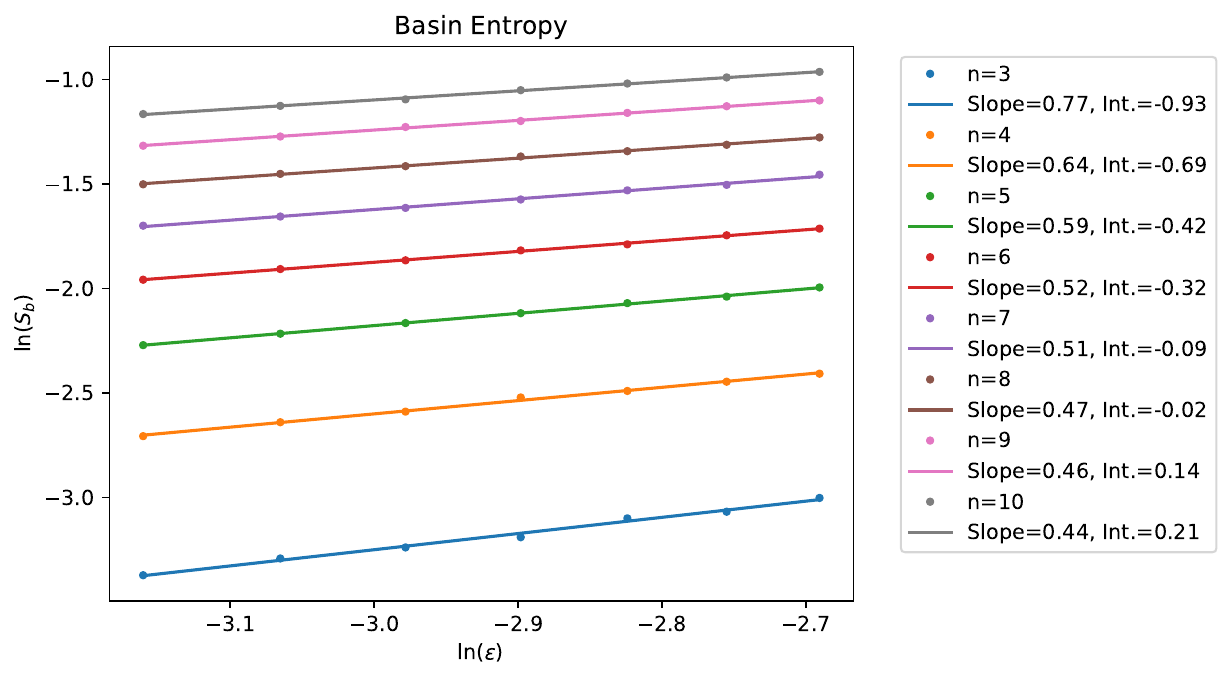}
        \caption{}
        \label{fig:basin_entropy}
    \end{subfigure}
    \begin{subfigure}{0.48\textwidth}
        \centering
        \includegraphics[height=0.22\textheight]{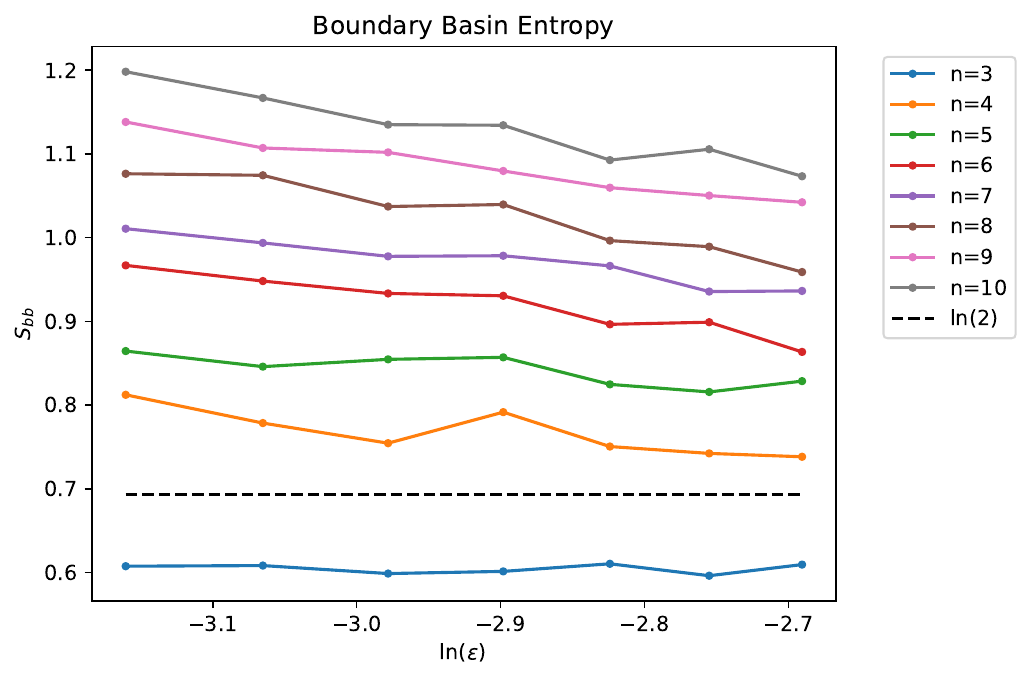}
        \caption{}
        \label{fig:boundary_basin_entropy}
    \end{subfigure}
    \caption{(a) Basin entropy by box length for $3\leq n\leq 10$. Points represent the calculated values, and the lines are linear regression in the log-log plot. The regression coefficients are shown next to the plot. (b) Boundary basin entropy by box length for $3\leq n\leq 10$. The broken line marks the value $\ln(2)$.}
    \label{fig:entropy}
\end{figure*}

Figure \ref{fig:basin_entropy} shows the basin entropy per $\varepsilon$ for the momentum escape basins with $n$ from $3$ to $10$. The plot also shows the linear regression in log-log space. The angular coefficient of each line gives the value of $\alpha$, and it follows the expectation, i.e., $\alpha$ decreases with increasing $n$. The linear coefficient is made up of a combination of both $A$ and $N_{\text A}$, and increases with $n$, also as expected. Overall, we can see from Figure \ref{fig:basin_entropy} that an increase of $n$ leads to a higher value of entropy for all the box scales $\varepsilon$.

Another useful quantity is the boundary basin entropy $S_{bb}$, which is similar to the basin entropy, but it is only applied to the boundary boxes. This quantity is defined as $S_{bb} = S/N_b$, where $S$ is calculated via \eqref{entropy} and $N_b$ is the total number of boundary boxes, i.e., boxes that contain more than one colour. In the supplementary information of paper \cite{daza2016basin}, the authors show that $S_{bb}>\ln(2)$ is a sufficient -- but not necessary -- condition for a boundary to be fractal. In Figure \ref{fig:boundary_basin_entropy}, we plot the boundary basin entropy for our system. We notice that for $n>3$ the boundary basin entropy is greater than $\ln(2)$ for all box sizes. Therefore, this demonstrates that all these boundaries are fractal. The case $n=3$, and only this case, has already been demonstrated to be fractal in the literature \cite{rod1973pathology, podolsky1998chaotic}.

\label{sec:entropy}
\section{Discussion}
In this paper, we have analysed the dynamics of geodesics in exact plane parallel gravitational waves given by equations \eqref{geodx} and \eqref{geody} or, in the Hamiltonian formulation, by equations \eqref{hamiltonian}-\eqref{potential}. This relativistic system is equivalent to the classical problem of the motion of free particles under a polynomial harmonic potential of degree $n$. The polynomial degree gives us a controlled family of models in which the number of escape channels is equal to this degree.

Our results show that the Wada property is robust with respect to variations in the polynomial degree, for both one-dimensional and two-dimensional escape basins, and persists across all cases considered. In this sense, pp-wave spacetimes provide a natural realization of open Hamiltonian systems with robust Wada basin topology. Furthermore, it is worth noting that in equation \eqref{potential}, we took the real part of the complex binomial expansion; nonetheless, our results are general and valid for the imaginary part by performing a $\pi/2n$ clockwise rotation \cite{brown2009complex}.

The system with $n=3$ is of particular interest, since the Hamiltonian for this case is a simpler version of the Hénon-Heiles system -- simpler since it does not have the quadratic terms in the potential. That is, it is a reduced system that still possesses the Wada property from the original Hénon-Heiles problem \cite{henon1964applicability,aguirre2001wada}. The dynamical system with harmonic polynomial potential of degree $n$ is also widely applicable to situations where the field of interest is Lagrange free, such as in Newtonian gravity, heat conduction, magnetostatics and others.

Furthermore, we introduced a quantitative description of unpredictability through basin entropy and boundary basin entropy. We showed that both measures are higher for increasing values of the polynomial degree $n$, as it leads to more exits and a higher boundary length. The slope of the lines of entropy versus grid length, as shown in Figure \ref{fig:basin_entropy}, also got shallower -- which indicates that the entropy remains high for all box length sides. The boundary basin entropy analysis showed that all the basins with $n>3$ are fractal. The $n=3$ basin, and only that one, was already known to be fractal in the literature \cite{rod1973pathology, podolsky1998chaotic}.

Taken together, these results establish a direct connection between relativistic geodesic dynamics and universal features of chaotic scattering, namely the coexistence of robust Wada basin topology and increasing entropy-based uncertainty in multi-exit systems.

\printbibliography

@article{aguirre2001wada,
  title = {Wada Basins and Chaotic Invariant Sets in the {{Hénon-Heiles}} System},
  author = {Aguirre, Jacobo and Vallejo, Juan C. and Sanjuán, Miguel A. F.},
  date = {2001-11-27},
  journaltitle = {Physical Review E},
  shortjournal = {Phys. Rev. E},
  volume = {64},
  number = {6},
  pages = {066208},
  issn = {1063-651X, 1095-3787},
  doi = {10.1103/PhysRevE.64.066208}
}

@article{bautista2021chaotic,
  title = {Chaotic Exits from a Weakly Magnetized {{Schwarzschild}} Black Hole},
  author = {Bautista, Joshua and Vega, Ian},
  date = {2021-08-05},
  journaltitle = {Classical and Quantum Gravity},
  shortjournal = {Class. Quantum Grav.},
  volume = {38},
  number = {15},
  pages = {155016},
  issn = {0264-9381, 1361-6382},
  doi = {10.1088/1361-6382/ac0e19}
}

@book{brown2009complex,
  title = {Complex Variables and Applications},
  author = {Brown, James Ward and Churchill, Ruel V.},
  date = {2009},
  series = {Brown and {{Churchill}} Series},
  edition = {8th ed},
  publisher = {McGraw-Hill Higher Education},
  location = {Boston},
  isbn = {978-0-07-305194-9},
  pagetotal = {468}
}

@article{daza2015testing,
  title = {Testing for {{Basins}} of {{Wada}}},
  author = {Daza, Alvar and Wagemakers, Alexandre and Sanjuán, Miguel A. F. and Yorke, James A.},
  date = {2015-12},
  journaltitle = {Scientific Reports},
  shortjournal = {Sci Rep},
  volume = {5},
  number = {1},
  pages = {16579},
  issn = {2045-2322},
  doi = {10.1038/srep16579}
}

@article{daza2016basin,
  title = {Basin Entropy: A New Tool to Analyze Uncertainty in Dynamical Systems},
  shorttitle = {Basin Entropy},
  author = {Daza, Alvar and Wagemakers, Alexandre and Georgeot, Bertrand and Guéry-Odelin, David and Sanjuán, Miguel A. F.},
  date = {2016-11},
  journaltitle = {Scientific Reports},
  shortjournal = {Sci Rep},
  volume = {6},
  number = {1},
  pages = {31416},
  issn = {2045-2322},
  doi = {10.1038/srep31416}
}

@article{daza2018wada,
  title = {Wada Structures in a Binary Black Hole System},
  author = {Daza, Álvar and Shipley, Jake O. and Dolan, Sam R. and Sanjuán, Miguel A. F.},
  date = {2018-10-29},
  journaltitle = {Physical Review D},
  shortjournal = {Phys. Rev. D},
  volume = {98},
  number = {8},
  pages = {084050},
  issn = {2470-0010, 2470-0029},
  doi = {10.1103/PhysRevD.98.084050}
}

@article{desouzafilho2020fractal,
  title = {Fractal Structures in the Deflection of Light by a Pair of {{Schwarzschild}} Black Holes},
  author = {De Souza Filho, E.E. and Mathias, A.C. and Caldas, I.L. and Viana, R.L.},
  date = {2020-12-28},
  journaltitle = {Indian Academy of Sciences Conference Series},
  shortjournal = {iascs},
  volume = {3},
  number = {1},
  doi = {10.29195/iascs.03.01.0006}
}

@article{dormand1980family,
  title = {A Family of Embedded {{Runge-Kutta}} Formulae},
  author = {Dormand, J.R. and Prince, P.J.},
  date = {1980-03},
  journaltitle = {Journal of Computational and Applied Mathematics},
  shortjournal = {Journal of Computational and Applied Mathematics},
  volume = {6},
  number = {1},
  pages = {19--26},
  issn = {03770427},
  doi = {10.1016/0771-050X(80)90013-3}
}

@article{fernandez2022weak,
  title = {Weak Dissipation Drives and Enhances {{Wada}} Basins in Three-Dimensional Chaotic Scattering},
  author = {Fernández, Diego S. and Seoane, Jesús M. and Sanjuán, Miguel A.F.},
  date = {2022-03},
  journaltitle = {Chaos, Solitons \& Fractals},
  shortjournal = {Chaos, Solitons \& Fractals},
  volume = {156},
  pages = {111891},
  issn = {09600779},
  doi = {10.1016/j.chaos.2022.111891}
}

@article{grebogi1983final,
  title = {Final State Sensitivity: {{An}} Obstruction to Predictability},
  shorttitle = {Final State Sensitivity},
  author = {Grebogi, Celso and McDonald, Steven W. and Ott, Edward and Yorke, James A.},
  date = {1983-12},
  journaltitle = {Physics Letters A},
  shortjournal = {Physics Letters A},
  volume = {99},
  number = {9},
  pages = {415--418},
  issn = {03759601},
  doi = {10.1016/0375-9601(83)90945-3}
}

@article{harris2020array,
  title = {Array Programming with {{NumPy}}},
  author = {Harris, Charles R. and Millman, K. Jarrod and family=Walt, given=Stéfan J., prefix=van der, useprefix=true and Gommers, Ralf and Virtanen, Pauli and Cournapeau, David and Wieser, Eric and Taylor, Julian and Berg, Sebastian and Smith, Nathaniel J. and Kern, Robert and Picus, Matti and Hoyer, Stephan and family=Kerkwijk, given=Marten H., prefix=van, useprefix=true and Brett, Matthew and Haldane, Allan and family=Río, given=Jaime Fernández, prefix=del, useprefix=true and Wiebe, Mark and Peterson, Pearu and Gérard-Marchant, Pierre and Sheppard, Kevin and Reddy, Tyler and Weckesser, Warren and Abbasi, Hameer and Gohlke, Christoph and Oliphant, Travis E.},
  date = {2020-09-17},
  journaltitle = {Nature},
  shortjournal = {Nature},
  volume = {585},
  number = {7825},
  pages = {357--362},
  issn = {0028-0836, 1476-4687},
  doi = {10.1038/s41586-020-2649-2}
}

@article{henon1964applicability,
  title = {The Applicability of the Third Integral of Motion: {{Some}} Numerical Experiments},
  shorttitle = {The Applicability of the Third Integral of Motion},
  author = {Henon, Michel and Heiles, Carl},
  date = {1964-02},
  journaltitle = {The Astronomical Journal},
  volume = {69},
  pages = {73},
  issn = {00046256},
  doi = {10.1086/109234}
}

@article{hunter2007matplotliba,
  title = {Matplotlib: {{A 2D Graphics Environment}}},
  shorttitle = {Matplotlib},
  author = {Hunter, John D.},
  date = {2007},
  journaltitle = {Computing in Science \& Engineering},
  shortjournal = {Comput. Sci. Eng.},
  volume = {9},
  number = {3},
  pages = {90--95},
  issn = {1521-9615},
  doi = {10.1109/MCSE.2007.55}
}

@article{kennedy1991basins,
  title = {Basins of {{Wada}}},
  author = {Kennedy, Judy and Yorke, James A.},
  date = {1991-08},
  journaltitle = {Physica D: Nonlinear Phenomena},
  shortjournal = {Physica D: Nonlinear Phenomena},
  volume = {51},
  number = {1--3},
  pages = {213--225},
  issn = {01672789},
  doi = {10.1016/0167-2789(91)90234-Z}
}

@article{meurer2017sympy,
  title = {{{SymPy}}: Symbolic Computing in {{Python}}},
  shorttitle = {{{SymPy}}},
  author = {Meurer, Aaron and Smith, Christopher P. and Paprocki, Mateusz and Čertík, Ondřej and Kirpichev, Sergey B. and Rocklin, Matthew and family=Kumar, given=AmiT, given-i={{Am}} and Ivanov, Sergiu and Moore, Jason K. and Singh, Sartaj and Rathnayake, Thilina and Vig, Sean and Granger, Brian E. and Muller, Richard P. and Bonazzi, Francesco and Gupta, Harsh and Vats, Shivam and Johansson, Fredrik and Pedregosa, Fabian and Curry, Matthew J. and Terrel, Andy R. and Roučka, Štěpán and Saboo, Ashutosh and Fernando, Isuru and Kulal, Sumith and Cimrman, Robert and Scopatz, Anthony},
  date = {2017-01-02},
  journaltitle = {PeerJ Computer Science},
  volume = {3},
  pages = {e103},
  issn = {2376-5992},
  doi = {10.7717/peerj-cs.103}
}

@article{podolsky1998chaos,
  title = {Chaos in Pp-Wave Spacetimes},
  author = {Podolský, Jiří and Veselý, Karel},
  date = {1998-09-01},
  journaltitle = {Physical Review D},
  shortjournal = {Phys. Rev. D},
  volume = {58},
  number = {8},
  pages = {081501},
  issn = {0556-2821, 1089-4918},
  doi = {10.1103/PhysRevD.58.081501}
}

@article{podolsky1998chaotic,
  title = {Chaotic Motion in Pp-Wave Spacetimes},
  author = {Podolský, J and Veselý, K},
  date = {1998-11-01},
  journaltitle = {Classical and Quantum Gravity},
  shortjournal = {Class. Quantum Grav.},
  volume = {15},
  number = {11},
  pages = {3505--3521},
  issn = {0264-9381, 1361-6382},
  doi = {10.1088/0264-9381/15/11/015}
}

@article{podolsky1999smearing,
  title = {Smearing of Chaos in Sandwich Pp-Waves},
  author = {Podolský, J and Veselý, K},
  date = {1999-11-01},
  journaltitle = {Classical and Quantum Gravity},
  shortjournal = {Class. Quantum Grav.},
  volume = {16},
  number = {11},
  pages = {3599--3618},
  issn = {0264-9381, 1361-6382},
  doi = {10.1088/0264-9381/16/11/310}
}

@article{poon1996wada,
  title = {Wada Basins Boundaries in Chaotic Scattering},
  author = {Poon, Leon and Campos, José and Ott, Edward and Grebogi, Celso},
  date = {1996-02},
  journaltitle = {International Journal of Bifurcation and Chaos},
  shortjournal = {Int. J. Bifurcation Chaos},
  volume = {06},
  number = {02},
  pages = {251--265},
  issn = {0218-1274, 1793-6551},
  doi = {10.1142/S0218127496000035}
}

@article{rod1973pathology,
  title = {Pathology of Invariant Sets in the Monkey Saddle},
  author = {Rod, David L.},
  date = {1973-07},
  journaltitle = {Journal of Differential Equations},
  shortjournal = {Journal of Differential Equations},
  volume = {14},
  number = {1},
  pages = {129--170},
  issn = {00220396},
  doi = {10.1016/0022-0396(73)90082-X}
}

@article{rossetto2020wada,
  title = {Wada {{Boundaries}} in Pp-{{Wave Spacetimes}}},
  author = {Rossetto, Pedro Henrique Barboza and Schelin, A.B.},
  date = {2020-12-28},
  journaltitle = {Indian Academy of Sciences Conference Series},
  shortjournal = {iascs},
  volume = {3},
  number = {1},
  doi = {10.29195/iascs.03.01.0004}
}

@book{stephani2003exact,
  title = {Exact {{Solutions}} of {{Einstein}}'s {{Field Equations}}},
  author = {Stephani, Hans and Kramer, Dietrich and MacCallum, Malcolm and Hoenselaers, Cornelius and Herlt, Eduard},
  date = {2003-03-27},
  edition = {2},
  publisher = {Cambridge University Press},
  doi = {10.1017/CBO9780511535185},
  isbn = {978-0-521-46702-5}
}

@article{virtanen2020scipy,
  title = {{{SciPy}} 1.0: Fundamental Algorithms for Scientific Computing in {{Python}}},
  shorttitle = {{{SciPy}} 1.0},
  author = {Virtanen, Pauli and Gommers, Ralf and Oliphant, Travis E. and Haberland, Matt and Reddy, Tyler and Cournapeau, David and Burovski, Evgeni and Peterson, Pearu and Weckesser, Warren and Bright, Jonathan and Van Der Walt, Stéfan J. and Brett, Matthew and Wilson, Joshua and Millman, K. Jarrod and Mayorov, Nikolay and Nelson, Andrew R. J. and Jones, Eric and Kern, Robert and Larson, Eric and Carey, C J and Polat, İlhan and Feng, Yu and Moore, Eric W. and VanderPlas, Jake and Laxalde, Denis and Perktold, Josef and Cimrman, Robert and Henriksen, Ian and Quintero, E. A. and Harris, Charles R. and Archibald, Anne M. and Ribeiro, Antônio H. and Pedregosa, Fabian and Van Mulbregt, Paul and {SciPy 1.0 Contributors} and Vijaykumar, Aditya and Bardelli, Alessandro Pietro and Rothberg, Alex and Hilboll, Andreas and Kloeckner, Andreas and Scopatz, Anthony and Lee, Antony and Rokem, Ariel and Woods, C. Nathan and Fulton, Chad and Masson, Charles and Häggström, Christian and Fitzgerald, Clark and Nicholson, David A. and Hagen, David R. and Pasechnik, Dmitrii V. and Olivetti, Emanuele and Martin, Eric and Wieser, Eric and Silva, Fabrice and Lenders, Felix and Wilhelm, Florian and Young, G. and Price, Gavin A. and Ingold, Gert-Ludwig and Allen, Gregory E. and Lee, Gregory R. and Audren, Hervé and Probst, Irvin and Dietrich, Jörg P. and Silterra, Jacob and Webber, James T and Slavič, Janko and Nothman, Joel and Buchner, Johannes and Kulick, Johannes and Schönberger, Johannes L. and De Miranda Cardoso, José Vinícius and Reimer, Joscha and Harrington, Joseph and Rodríguez, Juan Luis Cano and Nunez-Iglesias, Juan and Kuczynski, Justin and Tritz, Kevin and Thoma, Martin and Newville, Matthew and Kümmerer, Matthias and Bolingbroke, Maximilian and Tartre, Michael and Pak, Mikhail and Smith, Nathaniel J. and Nowaczyk, Nikolai and Shebanov, Nikolay and Pavlyk, Oleksandr and Brodtkorb, Per A. and Lee, Perry and McGibbon, Robert T. and Feldbauer, Roman and Lewis, Sam and Tygier, Sam and Sievert, Scott and Vigna, Sebastiano and Peterson, Stefan and More, Surhud and Pudlik, Tadeusz and Oshima, Takuya and Pingel, Thomas J. and Robitaille, Thomas P. and Spura, Thomas and Jones, Thouis R. and Cera, Tim and Leslie, Tim and Zito, Tiziano and Krauss, Tom and Upadhyay, Utkarsh and Halchenko, Yaroslav O. and Vázquez-Baeza, Yoshiki},
  date = {2020-03-02},
  journaltitle = {Nature Methods},
  shortjournal = {Nat Methods},
  volume = {17},
  number = {3},
  pages = {261--272},
  issn = {1548-7091, 1548-7105},
  doi = {10.1038/s41592-019-0686-2}
}

@article{yoneyama1917theory,
  title = {Theory of {{Continuous Set}} of {{Points}} (Not Finished),},
  author = {Yoneyama, Kunizo},
  date = {1917},
  journaltitle = {Tohoku Mathematical Journal, First Series},
  pages = {116}
}

@article{brinkmann1923riemann,
  title        = {On Riemann Spaces Conformal to Euclidean Space},
  author       = {Brinkmann, H. W.},
  date         = {1923-01},
  journaltitle = {Proceedings of the National Academy of Sciences of the United States of America},
  shortjournal = {Proc. Natl. Acad. Sci. U.S.A.},
  volume       = {9},
  number       = {1},
  pages        = {172--174},
  issn         = {0027-8424},
}

\appendix
\section{Numerical Methods}
\label{app_numerics}
The goal of this appendix is to give details about the numerical implementation of all the solution and analysis methods that would be otherwise distracting in the main body of the paper. The information presented here should be enough to guarantee that the reader can reproduce all the results exactly. Our particular implementation of the numerical algorithms was made in the \texttt{Python} programming language, using mostly the following libraries: \texttt{numpy}, for array calculations \cite{harris2020array}; \texttt{scipy}, for standard scientific tools \cite{virtanen2020scipy}; \texttt{matplotlib}, for plotting \cite{hunter2007matplotliba}; and \texttt{sympy}, for symbolic manipulation of expressions \cite{meurer2017sympy}. 

\subsection{Basin Calculations}

We solve the geodesic equations in their Cartesian form -- equations \eqref{geodx} and \eqref{geody} -- using the explicit Runge-Kutta method of order 5(4) as implemented by \texttt{Python}'s \texttt{Scipy} library \cite{dormand1980family,virtanen2020scipy}. For a particular value of $n$, the potential $V_n(x,y)$, equation \ref{potential}, can be calculated by a modified version of the binomial expansion formula. Nevertheless, we instead take advantage of \texttt{Python}'s \texttt{Sympy} library to generate the potential in a symbolic, programmatic manner \cite{meurer2017sympy}. We follow each particle until $x^2+y^2\geq 25$, then it is considered to have escaped through a particular channel.

To plot the escape basins displayed in figures \ref{fig:basin_n=5_1D}, \ref{fig:exit_basins} and \ref{fig:basin_n=5_zoom}, we solve the geodesic equations for several different initial conditions. For the one-dimensional basin (Figure \ref{fig:basin_n=5_1D}), we choose $p_{x,0} = p_{y,0} = 0$, $x_0 = \cos(\theta_0)$ and $y_0 = \sin(\theta_0)$, where $\theta_0$ varies from $0$ to $2\pi$ in an equally spaced array of $1000$ points. For the two-dimensional basins (figures \ref{fig:exit_basins} and \ref{fig:basin_n=5_zoom}), we choose $y_0 = 0$, $p_{x,0}$ and $p_{y,0}$ vary from $-1.5$ and $1.5$ in equally spaced arrays of $2000$ points, and $x_0 = (2-p_{x,0}^2-p_{y,0}^2)^{1/n}$. Values of initial momentum such that $p_{x,0}^2+p_{y,0}^2\geq 2$ were discarded since they lead to imaginary values for the initial position $x_0$. For the zoomed-in plots of Figure \ref{fig:basin_n=5_zoom}, we also used $2000$ points in each direction, with a magnification of $10$ in each zoom level, centred at $(p_x,p_y) = (0.7478,0.2854)$.

\subsection{Wada Check}

In Section \ref{sec:wada}, we have verified that the one-dimensional and two-dimensional escape basins have the Wada property for all $3\leq n \leq 10$. We have implemented the algorithm presented by \citeauthor{daza2015testing} in \citeyear{daza2015testing} \cite{daza2015testing} to test for the Wada property, which we refer to as the grid method. Further details of the algorithm can be found in the aforementioned paper.

The grid method works by refining (doubling the number of points) the regions neighbouring the boundary points. If, during refining, all colours are present, the algorithm proceeds to the next boundary point. If all boundary points are classified as Wada in this way, the algorithm halts. Nevertheless, we need a stop condition if not all colours are found, so the algorithm does not keep running indefinitely. For the one-dimensional basins, i.e. Table \ref{tab:wada_per_n_1D}, we choose the stop condition to be when the number of iterations is equal to $10$. For the two-dimensional basins, i.e. Table \ref{tab:wada_per_n_2D}, we choose the stop condition to be when no boundary point has been ``promoted'' to a boundary of more basins during the whole refinement process. These choices were made to minimize the exponentially growing computational cost while still achieving satisfactory results.

\subsection{Basin Entropy}

In Section \ref{sec:entropy}, we have calculated the basin entropy and the boundary basin entropy according to the definitions and algorithm presented by \citeauthor{daza2016basin} in \citeyear{daza2016basin} \cite{daza2016basin}. In this work, we used the two-dimensional basins with $2000\times2000$ resolution to compute the basin entropy. We have chosen values of $\varepsilon$ in a way to have from $30$ up to $50$ points, equally spaced, per covering box. The linear regressions displayed in Figure \ref{fig:basin_entropy} were calculated using the least squares polynomial fit method, as implemented in \texttt{Python}'s \texttt{numpy} library \cite{harris2020array}.

\end{multicols}
\end{document}